\documentclass{emulateapj}
\usepackage{amsmath}
\usepackage{amssymb}
\usepackage{graphicx}

\usepackage[hyperindex]{hyperref}
\hypersetup{
breaklinks = {false},
colorlinks = {false},
linkcolor={black},
pdfpagemode = {None}, 
pdfborder = {0 0 1},
pdftitle = {Measuring primordial non-Gaussianity through weak lensing peak counts},
pdfsubject = {},
pdfauthor = {Laura Marian, Stefan Hilbert, Robert E. Smith, Peter Schneider, Vincent Desjacques},
pdfkeywords = {gravitational lensing: weak, cosmological parameters, cosmology: primordial non-Gaussianity, cosmology: inflation, cosmology: observations, cosmology: theory, methods: numerical}
}

\def\sss{\scriptscriptstyle}

\def\ss{\scriptstyle}
\def\be{\begin{equation}}
\def\ee{\end{equation}}
\def\ba{\begin{eqnarray}}
\def\ea{\end{eqnarray}}
\def\Mpc{h^{-1}\,{\rm Mpc}}
\def\MpcI{h\,{\rm Mpc^{-1}}}
\def\kpc{h^{-1}\,{\rm kpc}}

\def\Msol{h^{-1}\,{\rm M_{\odot}}}
\def\deg2{\rm deg^2}
\def\arcmin2{\rm arcmin^2}
\def\fnl{f_{\rm NL}}
\newcommand{\euclid}{\textsc{euclid}}
\newcommand{\lsst}{\textsc{lsst}}
\newcommand{\mx}{\mbox}

\newcommand{\arcmint}{\mathrm{arcmin}}
\newcommand{\arcsect}{\mathrm{arcsec}}
\newcommand{\thetapix}{\theta_{\mathrm{pix}}}
\newcommand{\ngal}{\bar{n}_{\mathrm{gal}}}
\newcommand{\zmed}{z_{\mathrm{med}}}

\def\apjl{Astrophys.\ J.\ Lett.}
\def\mnras{Monthly Notices of the RAS}  

  \def\aap{Astron.\ Astrophys.}
\def\apj{Astrophys.\ J.}  
\def\apjs{Astrophys.\ J. Supp.}  
\def\prd{Phys.\ Rev.\ D} 

\def\jcap{Journal of Cosmology and Astro-Particle Physics}
\begin{document}
\title{Measuring primordial non-Gaussianity through weak lensing peak counts}
\author{Laura Marian$^{1}$, Stefan Hilbert$^{1}$, Robert
  E. Smith$^{2,1}$, Peter Schneider$^{1}$, Vincent
  Desjacques$^{2}$ \vspace{0.2cm}}
\affil{Argelander-Institut f\"{u}r Astronomie, Universit\"{a}t
  Bonn, Bonn, D-53121, Germany $^{1}$ \\ Institute for Theoretical
  Physics, University of Z\"{u}rich, Z\"{u}rich, CH 8057,
  Switzerland$^{2}$}
\email{lmarian@astro.uni-bonn.de}
\begin{abstract}
We explore the possibility of detecting primordial non-Gaussianity of
the local type using weak lensing peak counts. We measure the peak
abundance in sets of simulated weak lensing maps corresponding to
three models $\fnl={0, -100, 100}$. Using survey specifications
similar to those of \euclid{} and without assuming any knowledge of
the lens and source redshifts, we find the peak functions of the
non-Gaussian models with $\fnl=\pm100$ to differ by up to 15\% from
the Gaussian peak function at the high-mass end. For the assumed
survey parameters, the probability of fitting an $\fnl=0$ peak
function to the $\fnl=\pm 100$ peak functions is less than
0.1\%. Assuming the other cosmological parameters known, $\fnl$ can be
measured with an error $\Delta \fnl\approx 13$. It is therefore
possible that future weak lensing surveys like \euclid{} and \lsst{}
may detect primordial non-Gaussianity from the abundance of peak
counts, and provide complementary information to that obtained from
the cosmic microwave background.
\end{abstract}
\maketitle
\section{Introduction}

The inflationary paradigm is the leading theory of the early Universe,
of fundamental interest for cosmology and particle
physics. Understanding the mechanism and energy scale of inflation
remain major goals to attain, despite the continuous and fervent
efforts invested in this field.

A measurement of primordial gravitational waves would pin down the
energy scale of inflation, though it still belongs to the not-so-near
future. One possible way to discriminate between single- and
multi-field inflation models is to test the Gaussianity of the
primordial density fluctuations \cite{2004JCAP...10..006C}. The Cosmic
Microwave Background (CMB) has been so far the main and cleanest
inflationary probe. Recent results from the Wilkinson Microwave
Anisotropy Probe (WMAP) \cite{2010arXiv1001.4538K} suggested tentative
1$\sigma$-level evidence of primordial non-Gaussianity of the local
type, defined by the equation:
\be 
\Phi(\mx{\boldmath $x$})=\phi(\mx{\boldmath $x$})+\fnl[\phi^2(\mx{\boldmath
    $x$})-\langle \phi^2(\mx{\boldmath $x$})\rangle ].
\ee
The parameter $\fnl$ quantifies the local quadratic deviation of the
Bardeen potential $\Phi$ from a Gaussian potential $\phi$, and it is
currently constrained to the value $32\pm 21$
\cite{2010arXiv1001.4538K}.

It has long been suggested \cite{1986ApJ...310...19G,
  1988ApJ...330..535L, 1994ApJ...429...36F, 2000ApJ...541...10M} that
low-redshift observables can also be used to measure primordial
non-Gaussianity, despite the fact that the density field at such
redshifts is strongly non-Gaussian due to the action of gravity. In
the local non-Gaussianity models, there are mainly two effects on
low-redshift observables as recently outlined in the comprehensive
study of \cite{2008PhRvD..77l3514D}, and further explored in
\cite{2009MNRAS.396...85D, 2009MNRAS.398..321G, 2010MNRAS.402..191P,
  2009MNRAS.398.2143L, 2010arXiv1009.5085S}. First, $\fnl$ induces a
scale-dependence in the bias of dark matter halos, which affects
primarily the largest scales, i.e. $k<0.02 \,\MpcI$. Thus one can in
principle separate the $\fnl$ scale dependence from the gravitational
one, which occurs on smaller scales. Second, the abundance of massive
halos is higher/lower for positive/negative values of $\fnl$
\cite{2000ApJ...541...10M, 2008JCAP...04..014L}. As a final note, we
emphasize the importance of high- and low-redshift constraints on
non-Gaussianity as their comparison may provide insight into the
scale-dependence of $\fnl$ \cite{2010JCAP...09..026B,
  2010arXiv1010.3722S}.

In this paper we shall numerically investigate the sensitivity of weak
gravitational lensing (WL) peak counts to primordial non-Gaussianity
of the local type. The potential of WL surveys to constrain $\fnl$ has
already been tackled \cite{2004MNRAS.351..375A, 2010MNRAS.tmp.1684P,
  2010MNRAS.405..681F, 2010arXiv1010.0744O}, though without
considering shear peaks. Peak counts are a natural candidate for
$\fnl$ studies, since the largest of them are caused primarily by
massive halos. For WL studies of peak statistics, we mention the
works of \cite{2004MNRAS.350..893H, 2005ApJ...624...59H,
  2005NewA...10..676D, 2009ApJ...698L..33M, 2010PhRvD..81d3519K,
  2010MNRAS.402.1049D, 2010ApJ...709..286M} and the references
therein.  Should peak counts prove to be a sensitive $\fnl$ probe,
then one could easily use large future surveys like \euclid{}
\cite{2010arXiv1001.0061R} or the Large Synoptic Survey Telescope
(\lsst{}) \cite{2009arXiv0912.0201L} to obtain low-redshift
constraints on primordial non-Gaussianity.
\section{Method}
The observable that we use is the convergence field, i.e. the matter
density projected along the line of sight and scaled by a geometrical
factor. We study simulated WL convergence maps created from
ray-tracing through a suite of $N$-body simulations, generated with
the publicly available code $\ss\rm GADGET$
\cite{2005MNRAS.364.1105S}. A subset of these simulations was used and
described in the work of \cite{2009MNRAS.396...85D}. Three values of
$\fnl$ are considered: {0, -100, +100}, while the other cosmological
parameters are kept fixed. The cosmology matches the WMAP5 results
\cite{2009ApJS..180..330K}. We have a total of 18 simulations, with 6
realizations per $\fnl$ value. The initial conditions for each of the
6 sets of $\fnl$-model realizations are matched to reduce the cosmic
variance on the comparison of the peak functions corresponding to each
model. The box size is 1600 $\Mpc$, the number of particles is
$N=1024^3$, and the softening length is $l_{\rm soft}=40\, \kpc$.

We consider a survey similar to \euclid{} \cite{2010arXiv1001.0061R}
and to \lsst{} \cite{2009arXiv0912.0201L} for the WL simulations with:
an rms $\sigma_{\gamma}=0.3$ for the intrinsic image ellipticity, a
source number density $\ngal =40\,\arcmint^{-2}$, and a redshift
distribution of source galaxies given by ${\mathcal
  P}(z)=1.5\,z^2/z_0^3\,\exp[-(z/z_0)^{1.5}]$, where $z_0=0.6$. The
median redshift of this distribution is $\zmed=0.86$.

From each $N$-body simulation we generate 16 independent fields of
view. Each field has an area of $12 \times 12\,\deg2$ and is tiled by
$4096^2$ pixels, yielding an angular resolution $\thetapix
=10\,\arcsect$ and a total area of $\approx 14000\, \deg2$ for each
$\fnl$ model. The effective convergence $\kappa$ in each pixel is
calculated by tracing a light ray back through the simulation with a
Multiple-Lens-Plane ray-tracing algorithm \cite{2007MNRAS.382..121H,
  Hilbertetal2009}. Gaussian shape noise with variance
$\sigma_{\gamma}^2 /(\ngal\, \thetapix^{2})$ is then added to each
pixel, which creates a realistic noise level and correlation in the
filtered convergence field \cite{HilbertMetcalfWhite2007}.

For the peak finding we use an aperture filter
\cite{1996MNRAS.283..837S}, matching an Navarro-Frenk-White (NFW)
profile \cite{1997ApJ...490..493N} convolved with a Gaussian
function. Thus we adopt the convergence model: $\kappa_{\rm
  model}=\kappa_{\rm \sss NFW} \circ f_{\rm Gauss}$, where $f_{\rm
  Gauss}$ is a Gaussian function of width $f\times l_{\rm
  soft}$. $l_{\rm soft}$ is the softening length of the simulations,
$f=1.5$ for $M<7 \times 10^{14} \, \Msol$ and $f=2$ otherwise. Here
$\kappa_{\rm \sss NFW}$ is the NFW convergence profile truncated at
the virial radius, defined in \cite{2010ApJ...709..286M}. This model
agrees very well with the measured convergence profiles of the peaks
in the maps. It is useful both when working with simulations, since it
accounts for the lack of resolution below the softening scale, and
also when using real data, since shear data is difficult to obtain
near the centres of clusters.

The amplitude of the smoothed field at a point $\mx{\boldmath $x_0$}$ is
given by:
\be
\hat M(\mx{\boldmath $x_{0}$})=\int d^{2}x \, W(\mx{\boldmath $x_{0}$}-\mx{\boldmath $x$})
\kappa(\mx{\boldmath $x$}),
\label{eq:amp}
\ee
where $W$ is our filter and $\kappa$ is the convergence field. The
filter $W$ can be written as follows:
\be
W(x)={\mathcal C}_{\rm \sss W}\, \frac{\kappa_{\rm model}(x)
  -\bar \kappa_{\rm model}(R)}{\sigma_{\gamma}^2/\ngal},
\label{eq:fil1}
\ee
where ${\mathcal C}_{\rm \sss W}$ is a normalization constant and $R$
is the aperture radius, i.e. the radius over which the filter is
compensated, and $\bar\kappa(x)=2/x^2\int_{0}^{x}dy y \kappa(y)$. We
choose the normalization constant to be:
\be
{\mathcal C}_{\sss \rm W}= \frac{\sigma_{\gamma}^2}{\ngal}
\,\frac{M_{\rm \sss NFW}}{\int d^2x \,\kappa_{\rm model}^2(x)-\pi R^2\, 
{\bar \kappa_{\rm model}}^2(R)}\,.
\label{eq:norm}
\ee
If $\mx{\boldmath $x_{0}$}$ is the location of a peak created by an NFW
cluster of mass $M_{\rm \sss NFW}$ and redshift $z$, and the convergence
field is smoothed with a filter tuned to precisely such a cluster,
i.e. $\kappa_{\rm model}$ in Eq.~(\ref{eq:fil1}) corresponds to the
same $M_{\rm \sss NFW}$ and $z$, then this filter returns a maximum $S/N$
at $\mx{\boldmath $x_{0}$}$. At this location, the amplitude of the smoothed
map is:
\be
\hat M(\mx{\boldmath $x_{0}$}| M_{\rm \sss NFW}, z)=M_{\rm \sss NFW}\,.
\label{eq:assign}
\ee
The peak is assigned the mass $M_{\rm \sss NFW}$. If the peak does not
correspond to a real halo, and it is the result of line-of-sight
projections, then it can still be assigned an `effective' mass. In
practice, we smooth the convergence field with filters of various
masses, which yield different amplitudes (larger and smaller than the
filter mass) at the location of peaks. We interpolate these amplitudes
to determine the filter mass that would satisfy Eq.~(\ref{eq:assign}).

We choose $R$ to be the virial radius of the cluster to which the
filter is tuned: $R=(3M_{\rm \sss NFW}/800\,\pi\,\bar\rho)^{1/3}$,
where $\bar \rho$ is the mean density of the universe. Thus a filter
at a given redshift can be specified either through the mass $M_{\rm
  \sss NFW}$ or through its size $R$. We adopt the mass convention of
Sheth-Tormen \cite{1999MNRAS.308..119S}, with an overdensity defined
as 200 $\times$ the mean density (not the critical density), and the
concentration parameter of \cite{2008MNRAS.387..536G}. We evaluate the
$S/N$ of the mass estimator in a simplified scenario where we ignore
projection effects, and consider only the intrinsic ellipticity
noise. In this case the variance of the estimator is given by: ${\rm
  Var}(\hat M)=M_{\rm \sss NFW}\,{\mathcal C}_{\rm \sss W}$, and the
$S/N$ is obtained by combining this with the above equations. Note
that while the mass estimator does not depend on the shape noise due
to the normalization constant $\mathcal C$, the $S/N$ scales with it
as aperture filters usually do: $S/N \sim (\ngal/\sigma_{\gamma}^2)^{1/2}$.

\begin{figure*}
\centering{
\includegraphics[scale=0.75]{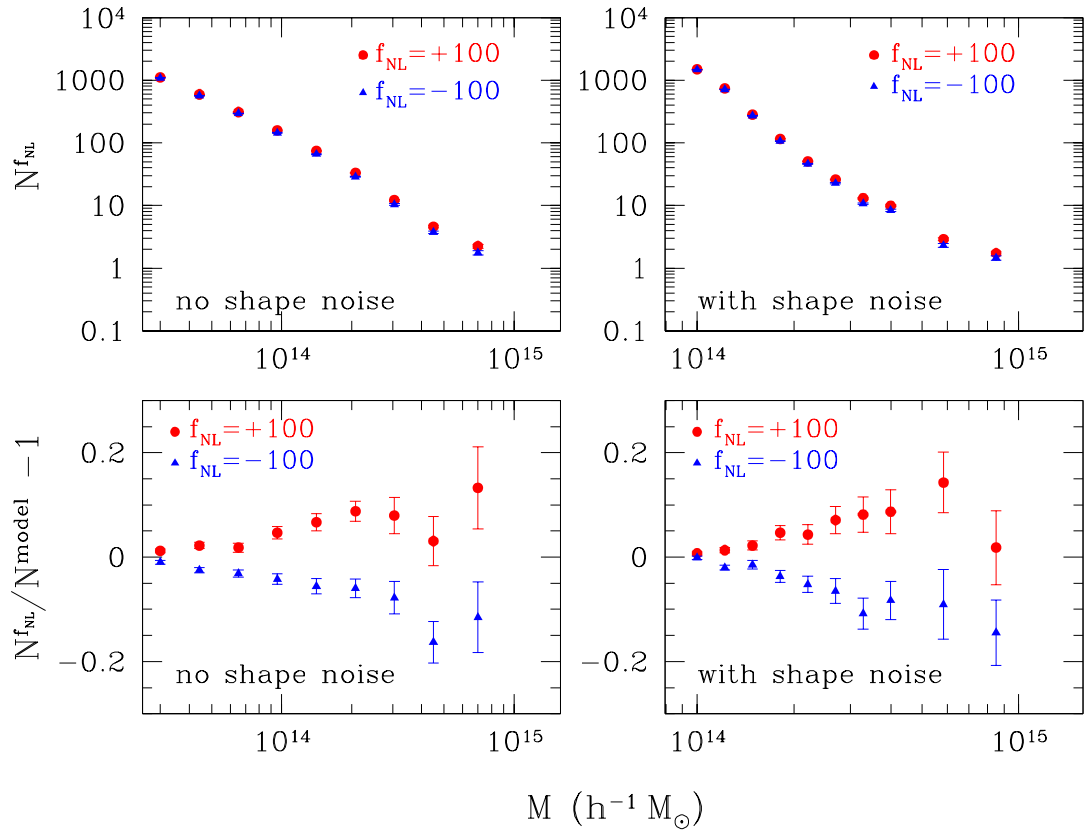}}
\caption {{\it Upper panels}: The measured peak functions for
  $\fnl=+100$ (red circles) and $\fnl=-100$ (blue triangles) for a
  survey with a median redshift $\zmed=0.9$ and an area of $14000
  \,\deg2$.  The filter used corresponds to redshift 0.3. The peak
  functions in the left panel are measured in the absence of shape
  noise, while in the right panel shape noise is included (assuming 40
  galaxies/$\arcmint^2$), and only the peaks with a $S/N > 3$ were
  selected (hence the smaller mass range on the $x$-axis).\vspace{0.5cm}
  \hspace{15cm}{\it Lower panels}: The fractional difference of the
  peak functions for the $\fnl= \pm 100$ models and the Gaussian model
  (red circles/blue triangles), without/with shape noise (left/right
  panel). The points are obtained as an average over all the fields of
  the fractional difference of the $\fnl \pm 100$ peak functions in
  every field and {\it the average} peak function in the Gaussian
  model.}
 
\label{fig:frac_diff}
\end{figure*}
Finally, in the case where no shape noise is included, we use the same
filter as in our previous work \cite{2010ApJ...709..286M}. This filter
can also be obtained by formally taking $\sigma_{\gamma}^2/\ngal=1$ in
Eqs~(\ref{eq:fil1}),~(\ref{eq:norm}) above.

The analysis of the convergence maps is carried out in two situations:
with and without shape noise. We adopt a very conservative approach in
which we consider known only the redshift distribution of the source
galaxies and the shape noise level, without any other information on
the sources or on the detected peaks. We also do not resort to
tomographic techniques. While this is an overly-pessimistic scenario
for a next-generation lensing survey like \euclid{} or \lsst{}, our
goal here is to provide a proof of concept of the possibility of using
WL peak counts to constrain primordial non-Gaussianity, rather than
the final quantitative answer to this question.

We perform a hierarchical smoothing of the maps with filters of
various sizes, from the largest down to the smallest, as described in
\cite{2009ApJ...698L..33M, 2010ApJ...709..286M}. This approach removes
the problem of `peaks-in-peaks' and it also naturally eliminates the
dependence of the measured peak function on a particular filter scale.
Since the median redshift of the source distribution is 0.9, we
adopt a fixed redshift of 0.3 for the matched filter described
above. For this redshift, the scale of the filter varies from
corresponding masses of $2\times 10^{15}\, \Msol$ to $3 \times 10^{13}
\,\Msol$ in the absence of shape noise, and $10^{14} \,\Msol$ in the
presence of it. The latter lower-limit choice of the filter scale is
due to the fact that shape noise contaminates seriously the smaller
peaks; imposing a minimum $S/N$ threshold alleviates but does not
remove the contamination. Peaks are detected in the smoothed maps as
local maxima, and are assigned a mass as described above.
\section{Results}
Figure~\ref{fig:frac_diff} presents the main result of this work. The
upper panels illustrate the peak functions measured in the $\fnl= \pm
100$ cosmologies, in the absence (left panel) and presence of shape
noise (right panel). For each $\fnl$ model, the points are the average
of the peak counts measured in all 96 fields and the error bars
represent errors on the mean. In the case of shape noise, we select
only peaks with a $S/N>3$. We note that the two average peak functions
are clearly distinct at the high-mass end, i.e. for $M>3\times
10^{14}\,\Msol$. This is better seen in the lower panels which show
the difference of the peak abundance measured in the $\fnl= \pm 100$
cosmologies, relative to the $\fnl=0$ peak abundance. We have
minimized the impact of the matched initial conditions of the
simulations on the error bars: the fractional difference is computed
as an average of the $\fnl \pm 100$ peak abundance in each field
ratioed to {\it the average} Gaussian peak abundance of 96 fields
(which represents our best model for the true Gaussian peak
abundance), as opposed to the average of the ratio of the non-Gaussian
and Gaussian peak functions measured in the same field. The latter
would have removed the cosmic variance of the fractional difference,
because of the matched initial conditions of the simulations from
which we built the convergence maps.

Just like in the case of the 3D halo mass function, (see for example
\cite{2009MNRAS.396...85D, 2010arXiv1009.5085S}), the peak functions
for the $\fnl$ models show a deviation from the Gaussian case. Unlike
the 3D studies which have presented halo mass functions measured at a
single redshift, the peak functions that we show here combine peaks in
the redshift range of the source distribution, and therefore are not
as regular and symmetric as their 3D counterparts. The asymmetry is
most likely due to modifications of the $S/N$ of peaks by
line-of-sight projections, and also by shape noise contamination. The
trend is similar however, with high-mass peaks displaying the largest
deviation. For both $\fnl$ models, this is about 10-15\% for the
largest mass bins i.e. $M>4 \times 10^{14}\,\Msol$.

To quantify the significance of the deviation, we perform a
$\chi^2-$test.  For the fiducial model $\fnl=0$, we estimate the
covariance of the counts in a field. We use the covariance of the mean
to obtain the $\chi^2$. For both $\fnl\pm 100$ we find a probability
$<0.1\%$ to fit the $\fnl\pm 100$ peak functions with an $\fnl=0$ peak
function. This is also true if we consider only the diagonal elements
(the variance of the mass bins) instead of the full covariance matrix,
and also if we vary the mass bins. We also use the measured counts to
estimate the Fisher error that a 14000 $\deg2$ WL survey would yield
on $\fnl$.  Assuming all other cosmological parameters known, the
forecasted error is $\Delta \fnl\approx 13$ for the fiducial value
$\fnl=0$. In the above we considered that the peak abundance scales
linearly with $\fnl$, similar to the 3D halo abundance. The values of
the $\chi^2$ and the Fisher error are largely maintained also if we
minimize the impact of the matched initial conditions of the
simulations, by using the first three simulations to compute the
$\fnl=+100$ peak function and the last three for the $\fnl=-100$
function.

Though these results are already very encouraging, it is possible to
improve measurements of primordial non-Gaussianity from WL surveys
even if one considers only peak counts. The most important is probably
the use of tomography, as numerical and analytical studies of the 3D
mass functions have shown that deviations of the halo abundance from
the Gaussian case significantly increase with redshift. Tomography
would allow peaks to be separated not only in terms of their
$S/N$--mass, but also of their redshifts, thus acquiring more
sensitivity to $\fnl$. On the other hand, one has to beware a possible
degeneracy with the amplitude of the matter power spectrum,
$\sigma_8$. This can be solved by either using as a prior very good
knowledge of $\sigma_{8}$ from other probes, such as the CMB
\cite{2006astro.ph..4069T}, WL, and large-scale structures or by
combining several observables sensitive to both $\fnl$ and $\sigma_8$,
as exemplified in \cite{2010arXiv1010.0744O}.
Our immediate goals are to further study how WL can be used to
constrain primordial non-Gaussianity, to build improved $\fnl$
estimators from WL observables, and to forecast $\fnl$ constraints
based on these observables.

For now we convey a simple, yet powerful statement: future WL surveys
could detect primordial non-Gaussianity of the local type from at
least one statistic--peak counts. In particular, surveys like
\euclid{} and \lsst{} should be able to provide complementary
information to the $\fnl$ information obtained from the ongoing CMB
mission Planck.\cite{2006astro.ph..4069T}.
\subsection*{Acknowledgements}
We kindly thank V. Springel for making public $\ss \rm GADGET$-2 and
for providing his B-FoF halo finder. LM, SH, and PS are supported by
the Deutsche Forschungsgemeinschaft (DFG) through the grant MA
4967/1-1, through the Priority Programme 1177 `Galaxy Evolution' (SCHN
342/6 and WH 6/3), and through the Transregio TR33 `The Dark
Universe'. RES and VD were partly supported by the Swiss National
Foundation under contract 200021-116696/1, the WCU grant R32-2008-
000-10130-0, and the UZ\"{u}rich under contract FK UZH 57184001. RES
also acknowledges support from a Marie Curie Reintegration Grant and
the Alexander von Humboldt Foundation.

\bibliographystyle{apj} 

\end{document}